# Network support of talented people[1]

Peter Csermely[2]

Department of Medical Chemistry, Semmelweis University, Budapest, Hungary

**Abstract** Network support is a key success factor for talented people. As an example, the Hungarian Talent Support Network involves close to 1500 Talent Points and more than 200,000 people. This network started the Hungarian Templeton Program identifying and helping 315 exceptional cognitive talents. This network is a part of the European Talent Support Network initiated by the European Council for High Ability involving more than 300 organizations in over 30 countries in Europe and extending in other continents. These networks are giving good examples that talented people often occupy a central, but highly dynamic position in social networks. The involvement of such 'creative nodes' in network-related decision making processes is vital, especially in novel environmental challenges. Such adaptive/learning responses characterize a large variety of complex systems from proteins, through brains to society. It is crucial for talent support programs to use these networking and learning processes to increase their efficiency further.

## Network Support of Talented People

Optimal development of creativity and talent needs the support of social networks (Csermely, 2013). As Eva Gyarmathy wrote recently: "Talent is the result of a highly efficient and active network-based functioning emerging in a network of diverse factors. Understanding and promoting it can best be achieved through a network of provision. Thus, gifted education and talent support itself should strive for a network-based structure" (Gyarmathy, 2016). Networking is a key success factor in a modern society (Christakis & Fowler, 2011; Csermely, 2009). Thus the acquisition of successful networking strategies is crucial for talented people. Safety seeking is a strategy, which makes a safety-net around the self, a tightly interwoven cluster of the family and best friends. The other networking strategy, novelty seeking builds on this safety-net. Novelty seeking requires the establishment of central, non-redundant social network contacts, which are filling a structural hole in social networks (Burt, 2005). This strategy needs a lifestyle, which is open for novelty and change. Talented people often treat novel situations as exciting challenges instead of conceiving them as a calamity, which they are unable to cope with. This is an important mindset for successful networking. Importantly, efficient networking strategies always keep in mind that networks show an extreme dynamism. There is an inner core of contacts (our family, our closest friends), which stays stable for many-many years. However, the outer shells of our social network structure continuously fade and recover. In the following part of my paper I will give examples of successful talent support networks.

---

## The Hungarian Talent Support Network

The Hungarian Talent Support Council (www.tehetseg.hu/en) is an umbrella organization organizing cooperation in talent support-related activities in Hungary and in the Hungarian speaking regions of Slovakia, Ukraine, Romania and Serbia (Csermely, 2013). There are close to 45 member organizations of the Council. These organizations support all types of talents, as examples those showing high ability in various sub-fields of science, crafts, innovation, sports or arts.

The Hungarian Parliament with a negligible number of opposing votes accepted a National Talent Program between 2008 and 2028 (http://tehetseg.hu/en/hungarian-talent-programme). In 2008 a National Talent Fund was also established having approximately 6 million USD per year. Half of this amount comes from income tax donations of approximately 300 thousand Hungarian citizens and the other half is doubling this amount from the Hungarian state budget. There are also large programs supported by the European Union, which are run by the Hungarian Talent Support Network. The first of these programs was the Hungarian Genius Program running between 2009 and 2012 having a budget of 1.8 million USD (http://tehetseg.hu/en/hungarian-genius-program). This program was followed by the Talent Bridges Program running between 2012 and 2015 having 8 million USD (http://tehetseg.hu/en/talent-bridges-programme-0). The organization started the Hungary of Talents Program in 2016, which will provide 25 million USD until 2020.

In the last ten years close to 1500 Talent Points (http://tehetseg.hu/en/talent-point-network) were established helping talented people to recognize and develop their abilities. These Talent Points are for example nurseries, schools, universities, but many of these Talent Points are carpentry shops, soccer clubs, chess clubs, and they even include penitentiaries. First and foremost, a Talent Point is running its own talent support program often including sub-programs in various areas of talents. However, besides this Talent Points are also information centers. A parent, teacher or friend may bring there a student or child, who shows signs of talent. Talented people may seek advice and help also themselves. Experts of the Talent Point (mostly highly trained psychologists in the field of talent identification and development) will help to assess the level of abilities, and to find adequate support.

The Hungarian Talent Support Network is a grass-root, self-organizing movement. Each week more and more new Talent Points ask for their registration. Close to 500 Talent Points (one third of total) are "accredited excellent" Talent Points meaning that they give professional help in a highly reliable and high quality manner. Talent Points form a network, and help each other discovering more than 35 thousand talents since 2010. Another important networking option is the self-organizing establishment of regional or thematic Talent Support Councils (http://tehetseg.hu/en/talent-support-councils). Talent Support Councils either organize local support of talented people, or promote the cooperation of whole areas, like that of mathematical, musical, Roma talents or talents requiring special need education. These close to a hundred local, thematic, or regional Talent Support Councils established their General Assembly in 2012. This grass-root movement efficiently mobilizes local resources, exchanges and spreads best practices, and formulates common needs and interests. Talent Points, local Talent Support Councils, talent support program, best practices and training programs are listed at the Talent Map (http://tehetseg.hu/tehetsegterkep) of Hungary and its neighboring countries.

Teachers play a central role in the identification of talented young people. From 2009 more than 23 thousand teachers were trained to be able to identify talented young people, and to be able to help them also using the nationwide support network. From 2006 there were several thousands Talent Days showing the local talent support possibilities for many interested participants and strengthening the local talent support community.

The various talent support programs developed several valuable resources. These resources include a library of freely accessible e-books (currently 40 issues: http://tehetseg.hu/tehetsegkonyvtar). These e-books cover aptitude and behavioral assessment, methods of mentoring, development of critical and reflective thinking, guidance on talent support in specialist fields, the role of talent care in integration, development of entrepreneurial and project management skills, etc. Three of the books are in English language. One contains interviews with 12 successful young talents (http://tehetseg.hu/konyv/driving-future-interviews-successful-young-talents, Proics and Jászay, 2015) and there are two English language books summarizing best practices of talent support from all around the world (http://tehetseg.hu/konyv/international-horizons-talent-support-i Gordon Győri, 2011; http://tehetseg.hu/konyv/international-horizons-talent-support-ii Gordon Győri, 2012). The whole library of 40 books is also available in printed forms.

**The Hungarian Templeton Program**
One of recent developments of the Hungarian talent support programs is the Hungarian Templeton Program (http://templetonprogram.hu/en). This program started in March 2015. The primary aim of the program is to develop and support exceptional Hungarian cognitive talents between ages of 10 and 29. After an intensive campaign mobilizing more than 20 thousand applicants by the intensive help of the Hungarian Talent Support Network (which had a much higher level of various abilities than the average of the population) 315 young Hungarians were identified with exceptional (1 out of 10,000) cognitive talents (http://templetonprogram.hu/en/our-fellows), and became Hungarian Junior Templeton Fellows participating in a one year intensive, personalized development program between March 2016 and March 2017.

The professional team of the Hungarian Templeton Program developed a new selection and talent support methodologies, which are designed to act as a sustainable model that we are happy to introduce to anyone interested. The goal of the program is to set up a creative community network of exceptionally talented people, who will , researchers and entrepreneurs over the next 10 to 30 years. It is also important to assist participants in using their talents in a responsible manner for the greater good of society, as well as to encourage society to better appreciate the enormous potential of these talented individuals and support the development of new talents. As a pilot study the program also provides assistance to 150 underprivileged children aged 5 to 8, who show exceptional cognitive talents.

The expert team of the Hungarian Templeton Program (involving Szilvia Péter Szarka, Éva Gyarmathy, Kristóf Kovács, supervisory psychologist of MENSA International and Benő Csapó, former vice-chair of the PISA Governing Board, among others) worked out a novel methodology for the on-line identification of exceptional cognitive talents from a large number of applicants identified and encouraged by the Hungarian Talent Support Network. Finally close to 20 thousand young people between 10 and 29 years registered to the Program. Applicants in the age group of 10 to 19 filled in 4 on-line tests measuring their fluent intelligence, verbal intelligence, working memory and problem solving abilities. Those 2800 young participants, who had the best results (where best results mean not only best average scores of the most successful 2 tests, but also exceptionally high scores in any of the 4 tests accomplished), had a second round of tests measuring their creativity and motivation, as well as submitting a reference letter. Applicants in the age group of 20 to 29 had to make an on-line fluent intelligence test, had to submit a CV focusing on internationally exceptional achievements, a motivation letter, a reference letter and (optionally) a motivational video or slide-show. The best 350 applicants of the two age groups were invited to personal interviews to judge the validity of their results, to fine-tune the final selection of the 315 Hungarian Junior Templeton Fellows and – last but not least – to draft the personal development program of the applicant – if selected as one of the participants in the Program. Those approximately 3000 young people, who showed rather outstanding achievements and results, but were not selected to the final 315 Hungarian Templeton Fellows, are helped by various personalized opportunities (such as scholarships, mentorships, etc.) offered by the Hungary of Talents Program.

The methodology of the personalized support program is tailored according to the individual needs of the 315 Fellows plus the 150 very young talents. Among other options there is individual, high quality mentoring (often involving coaching teams also helping to solve personal problems, increasing emotional maturity and resolving conflicts between Fellows and their environment), special e-learning and acceleration options, aiding international talent networking in various ways to help multi-social and intercultural experiences, development of a large number of soft skills (if necessary) such as communication skills, media-training, personal brand building, e-branding, as well as foreign languages, informatics etc. The program encourages the utilization of talent for the good of the society helping innovation, the development of financial and entrepreneurial skills (including start-up development), as well as social entrepreneurship (in cooperation with Ashoka Hungary) and noble purpose actions. As a key asset the program pairs Hungarian Templeton Fellows by the best and most recognized mentors. We have the partnership of the Hungarian Academy of Sciences and all the best Hungarian universities in this section of the program. The Hungarian Templeton Program is helped by an International Advisory Board. It is our great honor and privilege, that Joan Freeman, Jonathan Plucker and Rena Subotnik accepted the membership of this Board.

## The European Talent Support Network

The 2014 General Assembly of the European Council of High Ability (ECHA, www.echa.info) decided to support the formation of a European Talent Support Network. This Network
1. increases the identification and help of highly able young people in Europe;
2. boosts research activity in the field of high ability and help transfer findings to practice;
3. extends the current sharing of best practices in the field of high ability; and
4. demonstrates that people involved in the field of high ability have reached a "critical mass" at the European level, which needs to be taken into account when discussing EU and national policies in Europe (such as in education, research, innovation, social affairs, public health, etc.) related to high ability.

The European Talent Support Network consists of two types of participants: European Talent Centers and European Talent Points.

European Talent Centers organize activities in the field of high ability in a region or a whole country (there is a possibility for more than one European Talent Center per European country, e.g. in Austria and Germany we already have two Centers in Vienna and in Salzburg, as well as in Münster and Nürnberg-Regensburg, respectively). European Talent Centers are qualified by the Qualification Committee of ECHA led by Lianne Hoogeven. A European Talent Center
- has a well established expertise, activity and efficiency in coordinating talent support activity of a region or a country in Europe;
- has demonstrated paid or volunteer work devoted to talent support-related activities;
- volunteers or staff members of the Center have a proven expertise in research and/or practice of talent support;
- has the support of potential European Talent Points (see description later) representing at least 5 geographically distinct locations;
- has at least 5 internationally well-known experts in the field of high ability in form of an

International Advisory Board;
- able to provide high quality information on theoretical and practical issues of gifted education and talent support guiding the work of its European Talent Points, participating in discussions with other European Talent Centers and helping the formulation of proposals and joint actions at the European level;
- is willing and able to coordinate joint actions and other help of highly able young people (including e.g. regional, national, international events, like Talent Days, young musicians' competitions, young scientists competitions, Olympics of the Mind, etc.; development of qualifying curricula e.g. European Master Degree and research projects e.g. European Graduate School; education and training activities e.g. that of teachers, parents of highly able etc.);
- is able to keep records on the talent support activity of its region including the registration, help and coordination of European Talent Points and making this information available on the web (e.g. in form of a Talent Support Map of the region);
- is willing to cooperate with other European Talent Centers and with ECHA e.g.: in setting up joint actions, submitting joint proposals and grant applications, discussing joint policies and guidelines in the field of high ability, etc.;
- is open to be visited by representatives, experts, talented young people of other European Talent Centers;
- is willing to help and influence decisions on regional, national and/or European policies in the field of high ability (e.g. by public relations activities, media influence, lobbying etc.).

European Talent Points organize local activities in the field of high ability. European Talent Points can be
- organizations focusing mainly on talent support: research, identification, development of highly able young people (e. g: schools, university departments, talent centers, excellence centers, art, sport organizations focusing to talent development, NGOs, etc.);
- talent-related policy makers on national or international level (ministries, local authorities);
- business corporation with talent management programs (talent identification, corporate responsibility programs, creative climate);
- organizations of young people participating in talent support programs;
- organizations of parents of highly able children;
- or an umbrella organization (network) of the organizational types above.

The above types of organizations were only listed as examples. European Talent Centers may extend this list, but the organizations involved have to be related to the support of highly able young people. European Talent Points are registered by European Talent Centers using the following criteria. A European Talent Point
- has a strategy/action plan connected to talent (e.g. identification, various forms of support including complex programs, enrichment, competitions, etc., research, education, training, curriculum development, carrier planning, etc.) and a practice of this plan for minimum one year;
- is willing to share information on its talent support practices and other talent-related matters with other European Talent Points and European Talent Centers (by e.g. sharing programs, the strategy/action plan, data supporting its minimum one year of practice, best practices/research results on the web, organizing/attending joint conferences, organizing/attending joint Talent Days, etc.);
- is willing to cooperate with other European Talent Points including participation in joint programs, being open to be visited by representatives, experts, and/or talented young people of other European Talent Points.

The application process to become a European Talent Center was a widely publicized grass-root process. We have a great diversity of size, focus, type of expertise, type of activities and sources of support of European Talent Centers. This is an advantage, since in this way their strengths are complementing each other and they can also learn a lot from each other. So far we had two rounds of applications resulting in 19 European Talent Centers (http://echa.info/high-ability-in-europe/#). There will be future rounds of applications most probably each coming year, but definitely in 2017. Quite many of the applicants, not approved in one round, passes the selection threshold in one of the following rounds, since in the consultations before, during and after the selection process they realized the possibilities to improve their work and/or application further, and made already efforts to become qualified as a European Talent Center. Importantly, there is a growing interest to cooperate with the European Talent Support Network in other continents. This is why ECHA has established the status of Associated European Talent Centers and Associated European Talent Points in March 2016. The first two Associated Centers are from India and Peru. Associated Talent Points have been registered among others from Argentina, Brazil, Cuba, Dubai, Mexico and Saudi Arabia (http://echa.info/high-ability-in-europe/#associated-talent-centres). These organizations go through the same qualification process by using the same standards like their European counterparts with an important addition, that they must show evidence of regular and intensive previous cooperation with European organizations in the field of high ability.

What is the advantage of being qualified as a European Talent Center, and why is that more than administrative ballast? According to our initial experiences this qualification process increases several benefits of inter-organizational cooperation, such as

➢ exchange of best practices;
➢ increase the number of cross-country research projects in the field;
➢ increase of the visibility of our issues leading to a better chance to change gifted education policies;
➢ increase of the stability and robustness of everyday work (due to the increased exchangeability of colleagues in case of maternity/paternity-leave, sickness, personal problems, etc.);
➢ increase of community-feeling giving emotional and structural help for those participating in the network;
➢ increase of the effectiveness of using material resources in a region;
➢ increase of cooperation between talented young people enhancing their creative productivity (e.g. by using peer-pressure to become more excellent);
➢ extension of the number of gifted/talented people receiving recognition and support;
➢ extension of the number of people (teachers, mentors, parents, experts, scientists or business people) involved in talent support;
➢ creation of better and/or more effective chances to obtain local, corporate social responsibility, national and EU funding;
➢ establishment of internationally supported / grounded standards of talent management and talent support programs in a region, country or (finally) in Europe.

After the successful first European Youth Summit in March 2016 (http://www.youthsummit.eu/) the Network established a Youth Platform, which became a fast-growing group of talented young Europeans. In July 2016 Armin Fabian and Lukas Kyzlik were elected as a representative and deputy representative of the Youth Platform, respectively. Armin received the ''Magis Prize'', which is the biggest award of his school, given only for 2 students from more than 350. He won two times in a row the national round of the "Academic Competition of Biblical Studies" organized by the Romanian State. He received several diplomas from the County and City Council, plus a gold medal from the mayor of his city. He took part at 9 literature competitions, getting 4 international first prizes, 2 national rewards and one special gift. Lukas is a bachelor candidate at The Faculty of Electrical Engineering and Communication, Brno University of Technology. He works part-time as lector and creator of e-learning courses on electronics for gifted children at Intellectus since 2015. He is a member of the Talent Centre of the National Institute for Further Education (http://www.talentovani.cz/) since 2014. He took part in ASURO Robotics Course in 2014 and led student group conducting research on topic "Measurement of bodily functions using open-source electronics" in project T-Expedition 2015. He also took part in T-Study, a project allowing high school students to study selected university subjects, at Palacký University Olomouc in 2015-16.

The Network initiated several joint EU programs, such as an Erasmus+ strategic partnership program, a Creative Europe platform, as well as a Horizon2020 CO-CREATION program on "Education and skills: Empowering Europe's young innovators. The Network also plans to establish the European Talent Space, a 3D virtual world portal for talented young people in Europe. The European Talent Support Network gained the support of ~200 European Parliament members, the Commissioner of Education, Culture Youth and Sports, as well as Director Generals Martine Reicherts and Robert Jan Smits. Talent support became a priority of the Dutch, Slovakian and Maltese EU presidencies.

These wide-ranging activities show that the European Talent Support Network is a rapidly expanding self-organization process, mobilizing larger and larger amount of European people devoted to gifted education and talent support, and making high-quality work in their field to cooperate with each other in the various ways listed above. The Network is open to join any organizations willing to cooperate with its European counterparts. Those interested please write to the coordinator of the Network, Csilla Fuszek (http://www.echa.info/104-csilla-fuszek-s-cv) to qualification@echa.info or to the author of this paper.

## 'Creative nodes': An Often Occurring Social Network Position of Talented People

Creativity is not only characterized by the originality and usefulness of the product of the creative process. The level of creativity can also be predicted by the network position of the actor (Csermely, 2013). Considering various network positions three roles can be discriminated (Csermely, 2008; Farkas et al., 2011). Most of the network nodes are 'problem solvers'. These nodes are specialized to a certain task that they can perform with high efficiency. A few network nodes are so-called 'problem distributors'. These nodes are often hubs, thus have a larger number of network neighbors than the average. Problem distributors are specialized to the distribution of the responses to challenges already experienced by the network. Both problem solvers and problem distributors have a rather predictable behavior. Nodes of the third type exhibit a more unusual, unpredictable behavior. These 'creative nodes' are extremely dynamic, and continuously sample the entire network by frequent changes of their neighbors (Csermely, 2008).

In social networks the archetype of the creative node is the 'stranger' described by George Simmel more than a hundred years ago (Simmel, 1908). The stranger is different from anyone else. The stranger belongs to all groups, but at the same time does not belong to any of them. Malcolm Gladwell describes several 'stranger/creative nodes' in his book, "Tipping point" (Gladwell, 2002). These 'connectors' are interested in a large number of persons and information, which are different from each other. The wide and unbiased interest propels these boundary-spanning individuals to an integrative, central position in the social and information networks. Ronald Burt described a very similar network position as 'structural holes' (Burt, 2005). Neighbors of nodes occupying structural holes do not know each other. Moreover, not only the immediate neighbors of nodes occupying structural holes do not know each other, but the neighbors' second neighbors do not know each other either. Thus the creative node/stranger/connector/structural-hole network position is a great potential source of creativity. Such network positions are often occupied by talented people.

**The Role of 'Creative nodes' in Decision Making Processes of Complex Systems**

Alternating dominances of more plastic and more rigid complex system behaviors have been recently described as a general adaptation mechanism characterizing complex systems ranging from simple macromolecules to societies (Csermely, 2015). The core of this adaptation mechanism is that complex systems often display a bimodal distribution having either greater plasticity or greater rigidity. Plastic systems explore a large number of possible solutions, and thus, are highly adaptive to even unexpected changes of their environment. However, plastic systems do not have a 'memory', thus they can not reliably and efficiently produce the same optimal response to a repeated stimulus. On the contrary, rigid systems are highly optimized to a rather limited set of responses. Rigid systems are able to reproduce their limited set of responses reliably, but can not adapt to unexpected changes of their environment (Gáspár & Csermely, 2012; Csermely et al, 2013; Gyurkó et al, 2014; Csermely, 2015).

Increased human neuronal plasticity is characteristic to exploratory, creative periods (Ostby et al, 2012; Schlegel et al, 2013; Tagliazucchi et al, 2014). Positive emotions broaden the response repertoire (Fredrickson, 2004), increase the plasticity of the brain's mindset, and boost creativity. On the contrary, a rigid personality efficiently performs optimal solutions of previously practiced situations using previously fixed mental and behavioral sets displaying decisiveness and predictability. On one hand, extreme plasticity develops an inconsistent and undependable personality. On the other hand, extreme rigidity leads to stubborn behavior, which perceives ambiguous situations as 'threats' (Schultz & Searleman, 2002). Thus, optimal levels of creativity require alternating plasticity-dominated and rigidity-dominated mindsets.

Campbell's (Campbell, 1960) and Simonton's (Simonton, 1999) "blind-variation and selective retention model of creativity" is, in fact, describing the same plasticity-rigidity alterations that were described above. Creative thinking proceeds via shifts between generative and evaluative mindsets (Gabora, 2013; Sowden et al, 2015). Importantly, brainstorming involves separated plastic (idea-generating) and rigid (idea-selecting, idea-combining) segments (Osborn, 1953).

The social aspects of creativity (like community-aided idea selection) were emphasized by Mihály Csíkszentmihályi (Csíkszentmihályi, 1999). In the network context creativity is often displayed by special, creative nodes dynamically bridging a large number of distant network segments (Csermely, 2008; 2013). However, excess individual creativity can be detrimental to society, because creators invest in their unproven ideas at the expense of propagating proven ones (Gabora & Firouzi, 2013). Excess creativity is related to the "price of anarchy" in game theory (Roughgarden, 2005) showing the degradation of system's efficiency due to the selfish behavior of its agents. Importantly, many individuals can benefit from the creativity of the few without being creative themselves by copying creators (Nepusz & Vicsek, 2013; Gabora & Tseng, 2014). Finding the optimal level of group-creativity may require plasticity-rigidity alterations of social groups. In agreement with this assumption, an intermediate amount of long-range connections (Guimera et al, 2005; Uzzi & Spiro, 2005; Shore et al, 2015) resulting in the simultaneous presence of boundary spanning brokerage and trust-building closure (Tortoriello & Krackhardt, 2010; Uribe & Wang, 2014), as well as rotating leadership and contribution (Gloor et al, 2014) were shown to be key factors of team-success in business, arts, sports and science. Changes of group-plasticity and rigidity dominance may be an important learning mechanism of social groups as detailed below.

The organizational learning cycle (Dixon, 1994) has the same two major phases of (plastic) exploration and (rigid) selection, like the creative thinking process described above. The exploration/exploitation trade-off (March, 1991) representing the two phases of organizational learning exploring new possibilities and exploiting existing certainties, resembles to the plastic-rigid duality again. Changing dominance from exploration to exploitation was shown to be useful in early and late phases of firm development. Exploration and exploitation-dominance were termed as proactive and competitively aggressive orientation, respectively (Lumpkin & Dess, 2001). Exploration/exploitation dominance change was also shown in early and late phases of product development (Rothaermel & Deeds, 2004) and in plastic and rigid business environment (Lumpkin & Dess, 2001). Task switching (Monsell, 2003), the PDCA-cycle (plan-do-check-act/Shewhart/Deming-cycle; Shewhart, 1939; Deming, 1986) and the OODA-loop (Richards, 2004) are all plasticity-rigidity cycle variants helping decision making and process control. All these examples describe plasticity-rigidity alterations at the level of individuals and their social groups.

## The Potential Use of Complex System Adaptation and Learning Mechanisms in Talent Support

It will be an important future task of talent support practices (such as the talent support networks described in this contribution) to use the networking and learning processes described in the closing section of this contribution to increase the efficiency of talent support.

## Acknowledgments
The author thanks for the hundreds of thousands enthusiastic persons involved in Hungarian and European talent support, who participate in the talent support networks listed in this paper.

## Declaration of Conflicting Interests
The author declared no potential conflicts of interest with respect to the research, authorship, and/or publication of this article.

## Funding
The Hungarian Talent Support Network is supported by the Hungarian Talent Program and by the EU Structural Funds grant "Hungary of Talents' (EFOP-3.2.1-15). The Hungarian Templeton Program is supported by the Templeton World Charity Foundation (TWCF0117). The network-science related research was supported by the Hungarian National Science Foundation (OTKA K115378).

**Author Biography:** Peter Csermely is a professor of network science of the Semmelweis University (Budapest, Hungary; www.linkgroup.hu). In 1995 he launched an NGO providing research opportunities for >10,000 high school students. In 2006 he established the Hungarian Talent Support Council (www.tehetseg.hu/en) running a talent support network of ~200,000 people. From 2012 he is the president of ECHA (www.echa.info), which started the European Talent Support Network in 2015. He published 13 books and 270 research papers with citations over 12,000. Prof. Csermely is a member of the Hungarian Academy of Sciences and Academia Europaea, as well as an Ashoka, Fogarty, Howard Hughes, Rockefeller and Templeton Awardee.